\documentclass[preprint1]{emulateapj}
\usepackage{amsmath, amsbsy}
\usepackage{graphicx}
\usepackage{natbib, hyperref}
\begin{document}

% - front matter -
\title{Model-independent determination of curvature parameter by using $H(z)$ and $D_A(z)$ data pair from BAO measurement}

\author{Yun-Long Li\altaffilmark{1}, Shi-Yu Li\altaffilmark{2}, Tong-Jie Zhang\altaffilmark{2,3,4}, Ti-Pei Li\altaffilmark{1,5}}
\email{tjzhang@bnu.edu.cn}

\altaffiltext{1}{Department of Physics, Tsinghua University, Beijing 100084, China}
\altaffiltext{2}{Department of Astronomy, Beijing Normal University, Beijing 100875, China}
\altaffiltext{3}{Departments of Physics and Astronomy, University of California, Berkeley, CA 94720, USA}
\altaffiltext{4}{Lawrence Berkeley National Laboratory, 1 Cyclotron Road, Berkeley, CA 94720, USA}
\altaffiltext{5}{Key Laboratory of Particle Astrophysics, Institute of High Energy Physics, Chinese Academy of Sciences, Beijing, China}

\begin{abstract}
We present a model-independent determination of the curvature parameter $\Omega_k$ by using the Hubble parameter $H(z)$ and angular diameter distance $D_A(z)$ from the recent baryon acoustic oscillation (BAO) measurements. Each $H(z)$ and $D_A(z)$ pair from a BAO measurement can constrain a curvature parameter. The accuracy of the curvature measurement improves with increased redshift of $H(z)$ and $D_A(z)$ data. By using the $H(z)$ and $D_A(z)$ pair derived from BAO Lyman $\alpha$ forest measurement at $z=2.36$, the $\Omega_k$ is confined to be -0.05$\pm$0.06, which is consistent with the curvature $-0.037^{+0.044}_{-0.042}$ constrained by the nine-year WMAP data only. Considering future BAO meausurements, at least one order of magnitude improvement of this curvature measurement could be expected.
\end{abstract}

% - end of front matter -

% -
\section{INTRODUCTION}
The strong degeneracy between the curvature of the universe and the dark energy equation of state causes difficulties for constraining the two parameters simultaneously. The curvature is commonly left out in a dark energy analysis, or conversely, a dark energy constant is assumed in a determination of the curvature. However, a simple flatness assumption may result in erroneously reconstructing the dark energy equation of state even if the true curvature is very small \citep{Clarkson2007}, and a cosmological constant assumption may  arise confusions between a dynamical dark energy non-flat model and the flat $\Lambda$CDM model \citep{Virey2008}. In \citet{Clarkson2007}, when arguing the defects of a zero curvature assumption, a direct curvature determination by combining measurements of the Hubble parameter  $H(z)$ and the comoving angular diameter distance $D(z)$ was proposed:
\begin{equation}
 \Omega_k=\frac{[H(z)D'(z)]^2-c^2}{[H_0D(z)]^2}
 \label{eq:curv}
\end{equation}
where $'$ denotes derivative with respect to redshift $z$. This formula benefits from the Baryon Acoustic Oscillation (BAO) measurements which can provide $H(z)$ and $D_A(z)$ simultaneously at the same redshift. Here $D_A(z)$ is the angular diameter distance, which is correlated to the comoving angular diameter distance by $D(z)=(1+z)D_A(z)$. Since the derivation of $H(z)$ and $D_A(z)$ pairs from BAO measurements are purely geometrical, Eq.(\ref{eq:curv}) can be evaluated without any assumption of dynamic evolution of universe , therefore breaks the degeneracy between curvature and dark energy equation of state.

Several works have already derived the $H(z)$ and $D_A(z)$ pairs from the data of WiggleZ Dark Energy Survey at $z=0.44, 0.6, 0.73$ \citep{Blake2012}, and the third generation Sloan Digital Sky Survey (SDSS-III) at $z=0.35$ \citep{Chuang2012, Hemantha2013, Xu2013} and $z=0.57$ \citep{Reid2012, Kazin2013, Chuang2013, Anderson2014, Samushia2013}. With the quasar-Lyman $\alpha$ forest in SDSS-III, the measurement of $H(z)$ and $D_A(z)$ has extended to high redshift such as $z=2.36$ \citep{Font-Ribera2013}. These data can afford us to directly determine the curvature parameter via Eq.(\ref{eq:curv}).

The remaining issue is to estimate the $D'(z)$ reasonably. \citet{Mignone2008} has applied a novel method to take the derivative of the luminosity distance $D_L(a)$ with respect to the scale factor in their model-independent reconstruction of Hubble parameter. They decomposed the observables into a suitable basis functions, then recombined the derivatives of the basis functions to yield the $D'_L(a)$. The basis system was whereafter optimized by \citet{Maturi2009} to be capable to describe cosmologies independently of their background physics and improve the quality of the estimation of $D'_L(a)$. This method is independent to any cosmology model and can be employed by us to estimate the $D'(z)$.

By combining the data and method described above, we can determine the curvature parameter in a model-independent manner. This approach is different from previous works employing smoothing procedures in redshift bins or reconstruction of both Hubble parameter and comoving angular diameter distance \citep[e.g.][]{Mortsell2011}. The property of Eq.(\ref{eq:curv}) has determined that the error on the measured curvature parameter decreases as the redshift increases, thus we can benefit from the BAO Lyman $\alpha$ forest measurement which can provide $H(z)$ and $D_A(z)$ pair at high redshift.

The structure of this paper is as follows. In Section \ref{sec:method}, we review the essential parts of the model-independent method. The description of data and application of the method are shown in Section \ref{sec:application}. The discussions are presented in Section \ref{sec:discussion} and the conclusions are drawn in Section \ref{sec:conclusion}.

% -
\section{MODEL-INDEPENDENT METHOD}\label{sec:method}
We follow the idea in \citet{Clarkson2007}, and use the method proposed in \citet{Mignone2008} and further developed by \citet{Maturi2009, Benitez2012, Benitez2013}, to determine the curvature parameter in a model-independent manner.

\subsection{Estimating the $D'(z)$}
In Friedmann-Robertson-Walker metric, the comoving angular diameter distance is written as
\begin{equation}
%\begin{split}
D(z) =(1+z)D_A(z) =\frac{c}{H_0\sqrt{-\Omega_k}}\sin\left(\sqrt{-\Omega_k}\int_0^z{dz'\frac{H_0}{H(z')}}\right)
\label{eq:D(z)}
%\end{split}
\end{equation}
where the Hubble parameter $H(z)$ at late time is given by
\begin{equation}
\frac{H(z)}{H_0}=[\Omega_m(1+z)^3+\Omega_k(1+z)^2+\Omega_{\Lambda}F(z)]^\frac{1}{2}
\label{eq:H(z)}
\end{equation}
The parameters $\Omega_m$, $\Omega_k$ and $\Omega_{\Lambda}$ are the current density of matter, curvature and dark energy in unit of the critical density respectively.  The function
\begin{equation}
F(z)=\exp(3\int_0^z{\frac{1+w(z')}{1+z'}}dz')
\label{eq:F(z)}
\end{equation}
depends on the ratio $w(z)$ between the pressure and the density of the dark energy. By combining Eq.(\ref{eq:H(z)}) and Eq.(\ref{eq:F(z)}), the degeneracy between curvature and dark energy can be easily seen, as the non-zero curvature can be mimicked by the model with zero curvature and a dark energy component with $w(z)=-1/3$, therefore results in difficulties in fitting Eq.(\ref{eq:D(z)}) to observables. In fact, by solving the integral part from Eq.(\ref{eq:D(z)}) and taking derivative with respect to redshift on both side of the integral, one can derive Eq.(\ref{eq:curv}). The dynamic assumption of the universe thus has no place in the determination of the curvature.

The key issue here is to estimate $D'(z)$ --- the derivative of the comoving angular diameter distances with respect to the redshift in Eq.(\ref{eq:curv}). We suppose there is an underlying function $D(z)$ describing the behaviors of the comoving angular diameter distances. The $D(z)$ could be expanded into a series of suitable orthonormal functions $p_i(z)$,
\begin{equation}
  D(z)=\sum_{i=1}^M{c_ip_i(z)}
  \label{eq:decomp}
\end{equation}
By fitting Eq.(\ref{eq:decomp}) to the obsevables $D_{obs}$, the $M$ coefficients $c_i~(i=1,2,...,M)$ are determined. The number of the terms to be included in the expansion depends on the choice of the orthonormal basis and the quality of the data. The derivative of Eq.(\ref{eq:decomp}) is then taken as the estimation of the derivative in Eq.(\ref{eq:curv}).

The basis $\{p_i\}$, in principle, could be arbitrary with ideal data, but it is not in practice. \citet{Benitez2012} used an arbitrary orthonormal basis generated by Gram-Schmidt orthonormalization of $x^{i/2-1}~(i=0,1,...)$ to decompose the luminosity distances. They observed a systematic trend on the slope of the reconstructed $H(z)$ at intermediate redshifts when compared with the predictions of a $\Lambda$CDM cosmology, although they were consistent within the error bars. It indicated that a randomly chosen system of orthonormal basis functions may not be well adapted to the behavior of the measured data. \citet{Maturi2009} has proposed an optimal basis system derived from principal component analysis (PCA). Under this optimal basis system, the number of coefficients $M$ in Eq.(\ref{eq:decomp}) reaches minimal and the possible bias introduced by the choice of the basis is removed \citep{Benitez2013}.

% -
\subsection{Building the Optimal Basis}
The derivation of the optimal basis starts with writing the comoving angular diameter distances and their redshifts in column vectors $\boldsymbol{D}_{obs}$ and $\boldsymbol{z}$ respectively. The data set $\boldsymbol{D}_{obs}$ is therefore considered as a single point in an $n$-dimensional space, where $n$ is the number of the data points. Then we select a group of models that are believed to span the viable cosmologies, and calculate the comoving angular diameter distances $D(z)$ for each model at the redshifts in $\boldsymbol{z}$ to generate a set of vectors $\boldsymbol{D}_i$ $(i=1,2,...,M)$, where $M$ is the number of the models. These new vectors correspond to a cluster of points in the $n$-dimensional space and should meet the condition that the data set $\boldsymbol{D}_{obs}$ is tightly enclosed in the distribution of the cluster \citep{Benitez2013}. This ensemble of models, referred as the \textit{training set}, samples the possible behaviors of the comoving angular diameter distances, and initializes the PCA.

Once the training set models $\boldsymbol{T}_{n\times M}=(\boldsymbol{D}_1$
$, \boldsymbol{D}_2,...,\boldsymbol{D}_M)$ are defined, we build the so-called \textit{scatter matrix} $\boldsymbol{S=\Delta\Delta}^T$ with $\boldsymbol{\Delta}=(\boldsymbol{D}_1-\boldsymbol{D}_{ref}, \boldsymbol{D}_2-\boldsymbol{D}_{ref},...,\boldsymbol{D}_M-\boldsymbol{D}_{ref})$. $\boldsymbol{S}$ contains the differences between each training vector $\boldsymbol{D}_i$. $\boldsymbol{D}_{ref}$ is the reference model that defines the origin of the $n$-dimensional space. $\boldsymbol{\Delta}$ is actually the deviation of the training set from the reference model. $\boldsymbol{D}_{ref}$ could in principle be any combination of \{$\boldsymbol{D}_i$\}, and usually be set the average of the training set $\boldsymbol{D}_{ref}=\langle \boldsymbol{D}_i\rangle$. Any other choice of a reference model only affects the number of principal components (PCs) in decomposing the comoving angular diameter distances, e.g. $M$ in Eq.(\ref{eq:decomp}). It does not affect the final reconstruction of $D(z)$.

The PCs are derived by solving the eigenvalue problem $\boldsymbol{Sw}_i=\lambda_i\boldsymbol{w}_i$, where $\lambda_i$ and $\boldsymbol{w}_i$ are the $i$-th eigenvalue and eigenvector respectively. The eigenvector of the largest eigenvalue is the first PC. It corresponds to the direction on which the projection of $\boldsymbol{\Delta}$ has the largest variance, which means that the cluster of points mainly align in this direction. The second PC is the eigenvector of the second largest eigenvalue corresponding to the direction of the secondary largest variance, and so on. Since PCA aims at reducing the dimensionality of the training set substantially while retaining almost all the variation, an important issue arises that how many PCs should be employed in the reconstruction of $D(z)$. The selection criterion is based on the cumulative percentage of total variation \citep[section 6.1.1]{Benitez2013,Jolliffe2002} defined as:
\begin{equation}
  t_m=100\times \frac{\sum_{k=1}^m{\lambda_k}}{\sum_{k=1}^n{\lambda_k}}
\end{equation}
where $\lambda_k~(k=1,2,...,n)$ are the eigenvalues corresponding to the PCs and sorted in a descendent sequence $\lambda_k>\lambda_{k+1}$, $n$ is the number of total PCs, and $m$ is the number of PCs we shall employ in the reconstruction. The $t_m$, varying between 0 and 100, quantifies how much percentage of variance in the training set is preserved in the first $m$ PCs. After setting a threshold (e.g. $t_m>95$), $t_m$ returns the suitable number of PCs.

Then the deviation of the $D(z)$ from the reference model is decomposed by the $m$ PCs
\begin{equation}
  D(\boldsymbol{z})=\boldsymbol{D}_{ref}+\sum_{i=1}^m{c_i\boldsymbol{w}_i}
  \label{eq:decomp_pca}
\end{equation}
Eq.(\ref{eq:decomp_pca}) is similar with Eq.(\ref{eq:decomp}), except that the reference model here is not zero and the basis $\boldsymbol{w}_i$ is optimized via PCA. The coefficients $c_i$ are determined by fitting Eq.(\ref{eq:decomp_pca}) to $\boldsymbol{D}_{obs}$ through $\chi^2$ minimization.

The optimal basis for decomposition of $D(z)$ is also the optimal basis for decomposition of the $H_0$-independent angular comoving distances $\tilde{D}(z)=H_0D(z)$. When build the training set $\boldsymbol{\tilde{T}}_{n\times M}$ with $H_0$-independent angular comoving distances $\{\boldsymbol{\tilde{D}}_i\}$, Eq.(\ref{eq:decomp_pca}) can be rewritten as,
\begin{equation}
 D(\boldsymbol{z})=\frac{1}{H_0}\left(\boldsymbol{\tilde{D}}_{ref}+\sum_{i=1}^m{\tilde{c_i}\boldsymbol{w}_i}\right)
  \label{eq:decomp_pca2}
\end{equation}
where the $\boldsymbol{\tilde{D}}_{ref}$ is the $H_0$-independent reference model. In the $\chi^2$ minimization step, we use Eq.(\ref{eq:decomp_pca2}) to leave the Hubble constant $H_0$ as a free parameter. The $D'(z)$ in Eq.(\ref{eq:curv}) is derived by taking derivative on both side of Eq.(\ref{eq:decomp_pca2}) with respect to redshift.

% -
\section{APPLICATION TO REAL DATA}\label{sec:application}

\subsection{The Data}
The $H(z)$ and $D_A(z)$ pairs are collected from current literatures of BAO measurements. Since the SDSS and  BOSS CMASS samples have been analyzed multiple times, we tend to use the most recent published results to avoid overlap. The $(H,D_A)$ pairs and the BAO surveys are listed in Table \ref{tab:BAO}. We consider the covariance between distances constrained by different BAO samples at different redshift ranges to have negligible effect on our determination of derivative of distance with respect to redshift.

\begin{deluxetable}{llll}
%\tabletypesize{\small}
\tablecaption{Hubble parameter $H(z)$ and angular diameter distance $D_A(z)$ at the same redshift $z$ taken from BAO measurements\label{tab:BAO}}
\tablewidth{0pt}
\tablehead{
\colhead{$z$} & \colhead{$H(z)$} & \colhead{$D_A(z)$} & \colhead{Survey}\\
\colhead{}   & \colhead{(km~s$^{-1}$~Mpc$^{-1}$)} & \colhead{(Mpc)} & \colhead{}
}
\startdata
 0.44 & $82.6\pm7.8$ & $1205\pm114$ & WiggleZ \tablenotemark{a} \\
 0.6  & $87.9\pm6.1$ & $1380\pm95$  &          \\
 0.73 & $97.3\pm7.0$ & $1534\pm107$ &          \\ \hline
 0.35 & $84.4\pm7.0$ & $1050\pm38$  & SDSS DR7 \tablenotemark{b} \\ \hline
 0.57 & $93.1\pm3.0$ & $1380\pm23$  & BOSS DR11 CMASS \tablenotemark{c} \\ \hline
 2.36 & $226\pm8$    & $1590\pm60$  & BOSS DR11 Ly-$\alpha$ forest \tablenotemark{d}
\enddata
\tablenotetext{a}{\citet{Blake2012}}
\tablenotetext{b}{\citet{Xu2013}}
\tablenotetext{c}{\citet{Samushia2013}}
\tablenotetext{d}{\citet{Font-Ribera2013} }
\end{deluxetable}

To lower the statistical errors in $D'(z)$ estimation, we enlarge the distance samples with the luminosity distances $D_L(z)$ from the currently largest homogeneously reduced compilation of SN Ia, the Union2.1 \citep{Suzuki2012}, which contains 580 SNe Ia. The data are in form of distance modulus $\mu$ which should be converted to the luminosity distance by
\begin{equation}
 D_L=10^{\frac{\mu}{5}-5}
\end{equation}
The corresponding comoving angular diameter distances are derived via the Etherington's relation which holds for any space-time $D(z)=D_L(z)/(1+z)$. For simplicity, we use the covariance matrix of distance modules of SNe Ia with statistical errors only, which is in diagonal form, during the $\chi^2$ minimization procedure.

% -
\subsection{Results}
\begin{figure}
  \epsscale{1.0}
  \plotone{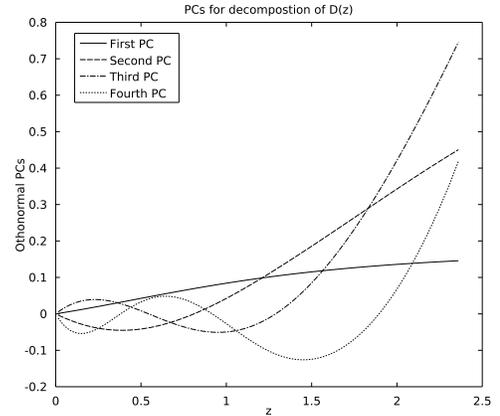}
  \caption{First four principal components derived from non-flat $\Lambda CMD$ models with the $\Omega_m$, $\Omega_{\Lambda}$ and $\Omega_k$ sampling the ranges $0.1<\Omega_m<0.5, 0.5<\Omega_{\Lambda}<0.9$ and $-0.2<\Omega_k<0.2$ one thousand times. The reference model takes the mean vector of the 1000 training vectors.}
  \label{fig:PCs}
\end{figure}

Fig.\ref{fig:PCs} has depicted the first four PCs for the comoving angular diameter distances. The training set $\boldsymbol{\tilde{T}}_{n\times M}$ is built from 1000 non-flat $\Lambda$CDM  models with the parameters uniformly sampled in the cubic parameter space with boundaries $0.1<\Omega_m<0.5,~0.5<\Omega_{\Lambda}<0.9$ and $-0.2<\Omega_k<0.2$, the sum of $\Omega_m$, $\Omega_{\Lambda}$ and $\Omega_k$ can deviate from 1. Here the reference model takes the average of the training set $\boldsymbol{\tilde{D}}_{ref}=\langle \boldsymbol{\tilde{D}}_i \rangle$. A different reference model would not change the reconstruction of $D(z)$ as long as the distribution of the training set tightly enclosing the data \citep[][section 3]{Benitez2013}.
The first PC retains 98.8\% of the total variance in the sample, which means it has already considered the major properties in the expansion of the training set. We therefore use the first PC to decompose the $D(z)$. Here we emphasis that, although the PCs are determined by the training set sampled from different non-flat $\Lambda$CDM models, they are able to constrain other cosmologies that are not explicitly contained in the training set \citep[see][section 4.2]{Maturi2009}.

\begin{figure}
  \epsscale{1.0}
  \plotone{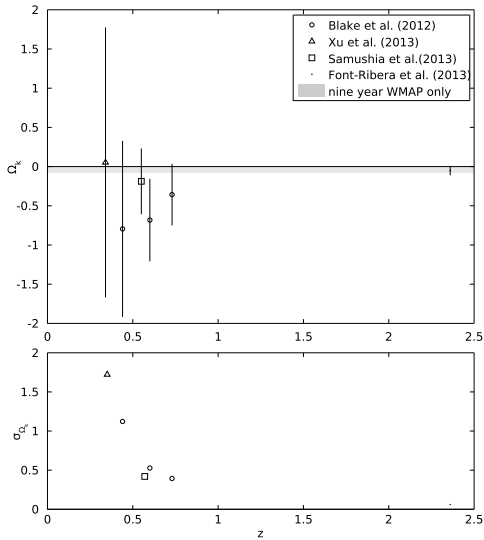}
  \caption{Upper panel: Curvatures measured by Eq.(\ref{eq:curv}) using $H(z)$ and $D_A(z)$ pairs from BAO measurements. Each $(H,D_A)$ pair constrains a curvature parameter. For comparison, the gray region marks the curvature constrained by the nine year WMAP data only \citep{Bennett2013}. Lower panel: $1\sigma$ errors in this curvature measurements. }
  \label{fig:curv}
\end{figure}

Fitting Eq.(\ref{eq:decomp_pca2}) to the observables $\boldsymbol{D}_{obs}$ yields the coefficients $c_1=(-11.3\pm9.2)\times 10^4$ and $H_0=69.8\pm1.9$~km s$^{-1}$ Mpc$^{-1}$. By substituting the coefficients and BAO data into Eq.(\ref{eq:decomp_pca2}) and Eq.(\ref{eq:curv}), each $(H,D_A)$ pair has derived a curvature parameter.  The upper panel of Fig.\ref{fig:curv} has shown the $\Omega_k$ with $1\sigma$ errors. The curvature measurement using high redshift data at $z=2.36$ has the best constrain that $\Omega_k=-0.05\pm0.06$, which is consistent with the curvature $\Omega_k=-0.037^{+0.044}_{-0.042}$ constrained by the nine year WMAP data only \citep{Bennett2013}. At low and medium redshifts, the deviations of $\Omega_k$ from zero are although nearly of order unit, the measurements are still consistent with a flat universe.

The lower panel of Fig.\ref{fig:curv} has shown a trend that the errors in $\Omega_k$ decrease with increasing redshift. By comparing the errors of the input $\left(H,D_A\right)$ data with the errors of the output $\Omega_k$, we have noticed that even if the BAO data at higher redshift may have less precisions than the lower ones, the curvatures determined by higher redshift BAO data still have smaller errors, such as 3.7\% for $D_A(2.36)$ and 3.5\% for $H(2.36)$ while 1.6\% for $D_A(0.57)$ and 3.2\% for $H(0.57)$.

\section{DISCUSSIONS}\label{sec:discussion}
Consider the error propagation formula of Eq.(\ref{eq:curv})
\begin{equation}
\begin{split}
&(\Delta\Omega_k)^2 = \sum_{\begin{subarray}{c}\alpha \in \{H,H_0,\\D,D'\}\end{subarray}}{\left(\frac{\partial\Omega_k}{\partial \alpha}\right)^2(\Delta \alpha)^2}\\
&=4\left[\Omega_k+\frac{c^2}{(H_0D)^2}\right]^2\left[\left(\frac{\Delta H}{H}\right)^2+\left(\frac{\Delta D'}{D'}\right)^2\right]\\
& ~+ 4\Omega_k^2\left[\left(\frac{\Delta D}{D}\right)^2+\left(\frac{\Delta H_0}{H_0}\right)^2\right]
\end{split}
\label{eq:errors}
\end{equation}

We can find that the behavior of $\sigma_{\Omega_k}^2$ is dominated by $D(z)^{-4}$ at low redshift. This feature results in large errors in the determination of curvature when using low redshift data. Nevertheless, Eq.(\ref{eq:errors}) tends to be a constant which only depends on the errors of $H,H_0,D,D'$ when using very high redshift data, revealing that Eq.(\ref{eq:curv}) could tightly constrain the curvature parameter via precise measurements of $H(z)$ and $D_A(z)$ at high redshifts.

We repeat the procedure in previous sections to a synthetic sample simulated by a standard flat $\Lambda$CDM model with parameters $\Omega_m=0.315$, $\Omega_{\Lambda}=0.685$, $h=0.673$ from \citet{Planck2013}. The uncertainties of $H(z)$ and $D(z)$ measurements resemble a theoretical BAO cosmic variance forecast for a full-sky BAO survey in Table 2 of \citet{Weinberg2012} , in which case the $H(z)$ and $D_A(z)$ measurements can reach a precision of 0.2\% at $z>1$.

\begin{figure}
  \epsscale{1.0}
  \plotone{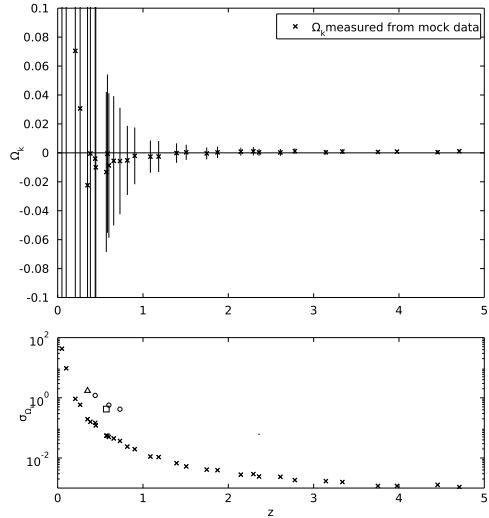}\\
  \caption{Upper panel: the curvature parameter measured from a synthetic sample generated by a flat $\Lambda$CDM model. Lower panel: 1$\sigma$ errors of the curvatures. The cross are curvature errors measured from mock data. The circles, triangle, square and dot are from real BAO data and the same with the dots in the lower panel of Fig.\ref{fig:curv}}
  \label{fig:curv_mock}
\end{figure}

Fig.\ref{fig:curv_mock} has shown the results of the curvature measurements on the mock data. The tendency that $\sigma_{\Omega_k}$ changes with $z$ is in agreement with those derived from real BAO data. With precise BAO measurements in the future, we could expect this method to constrain the curvature parameter within $10^{-3}$ error limit or better.

\section{CONCLUSIONS}\label{sec:conclusion}
Based on the work of \citet{Clarkson2007}, \citet{Mignone2008}, \citet{Maturi2009} and \citet{Benitez2013}, we present a model-independent method to determine the curvature parameter. The $H(z)$ and $D_A(z)$ pairs involved in this method are derived by BAO measurements which only depend on the space-time geometry, thus can afford us to measure the curvature without any assumptions of the dynamic evolution of the universe. The luminosity distances $D_L(z)$ from Union2.1 SN Ia compilation are included to have a better constrain on estimation of the derivative of comoving angular diameter distance with respect to redshift. The curvature parameters measured in this work are in agreement of a flat universe within error limits.

The feature of Eq.(\ref{eq:curv}) leads to the fact that the accuracy of the curvature measurement improves with increasing redshift of $H$ and $D_A$ and will reach a limit primarily determined by the data quality. In this work, the errors of curvature measurements at low redshift are nearly of order unit, while the curvature measurement at a high redshift $z=2.36$ has derived a much better constrain $\Omega_k=-0.05\pm0.06$, which is consistent with the nine year WMAP-only results.

We use the density parameters from \citet{Planck2013} and theoretical BAO cosmic variance forecast from \citet{Weinberg2012} to generate a small synthetic sample to test the curvature measurement. At least one order of magnitude improvement of this curvature measurement could be expected.

\acknowledgments
This work is supported by the National Science
Foundation of China (Grant No. 11033003), the National Science
Foundation of China (Grant No. 11173006), the Ministry
of Science and Technology National Basic Science program
(project 973) under Grant No. 2012CB821804.

\textit{Note added}---We were aware of \citet{Sapone2014} who have accomplished the similar work during our preparation of this paper.

%\bibliography{theBib}

\end{document}